\documentclass[pre,twocolumn,superscriptaddress,showpacs]{revtex4}
\usepackage{amsmath,amssymb}
\usepackage[utf8x]{inputenc}
\usepackage[T1]{fontenc}
\usepackage[dvipdfm]{graphicx}
\usepackage[english]{babel}

\begin{document}
\title{Endemic infections are always possible on regular networks}
\author{Charo I. \surname{Del Genio}}\email{C.I.del-Genio@warwick.ac.uk}
	\affiliation{Warwick Mathematics Institute, University of Warwick, Gibbet Hill Road, Coventry CV4 7AL, UK}
	\affiliation{Centre for Complexity Science, University of Warwick, Gibbet Hill Road, Coventry CV4 7AL, UK}
	\affiliation{Warwick Infectious Disease Epidemiology Research (WIDER) Centre, University of Warwick, Gibbet Hill Road, Coventry CV4 7AL, UK}
	\affiliation{\mbox{Max Planck Institute for the Physics of Complex Systems, Nöthnitzer Str. 38, Dresden D-01187, Germany}}
\author{Thomas \surname{House}}\email{T.A.House@warwick.ac.uk}
	\affiliation{Warwick Mathematics Institute, University of Warwick, Gibbet Hill Road, Coventry CV4 7AL, UK}
	\affiliation{Centre for Complexity Science, University of Warwick, Gibbet Hill Road, Coventry CV4 7AL, UK}
	\affiliation{Warwick Infectious Disease Epidemiology Research (WIDER) Centre, University of Warwick, Gibbet Hill Road, Coventry CV4 7AL, UK}

\date{\today}

\begin{abstract}
We study the dependence of the largest component in regular networks
on the clustering coefficient, showing that its size changes smoothly
without undergoing a phase transition. We explain this behaviour via
an analytical approach based on the network structure, and provide an
exact equation describing the numerical results. Our work indicates
that intrinsic structural properties always allow the spread of epidemics
on regular networks.
\end{abstract}


\maketitle

Many natural and engineered systems can be easily represented as networks,
where discrete elements, the nodes, interact via a set of links. The use of a
network paradigm has proved particularly helpful in mathematical epidemiology,
allowing substantial advances in our understanding of the spread of
diseases~\cite{Alb02,Boc06,Dan11,Pas01}.  A common approach, especially in
ecological or spatially-embedded epidemiological models, is to use regular
networks, in which all the nodes have the same number of links
$k$~\cite{Shi05,Lel09,DurXX,KeeXX,Giv11,Abd11}.  Specific examples of disease
applications include systems as diverse as bubonic plague~\cite{Kee00},
foot-and-mouth disease~\cite{Fer01}, and citrus tristeza virus~\cite{Gib97},
each of which requires a different regular network topology.  As $k$ is fixed,
the number of links in a regular network is proportional to the number of nodes
and regular networks belong to the group of sparse networks, which encompass
several systems of general interest~\cite{Del11}. An important feature of
networks that influences their structure as well as the dynamics of the
processes they support is the clustering coefficient $c$, defined as the number
of closed triplets divided by the total number of triplets. Multiple
experimental studies have evidenced how real-world networks often have large
values of $c$, and specific random graph models have been formulated to
reproduce the observed clustering
properties~\cite{Wat98,NewXX,Vol04,Gle09,Bol11}. The recognized importance of
the clustering coefficient in epidemiological dynamics has caused it to be used
as a control parameter in many works~\cite{BriXX,Mos09}, including on regular
networks~\cite{Mol12}. However, despite much recent progress, there is still no
unified understanding of the full impact of clustering on networks. For large
enough clustering coefficients, sparse networks such as the regular ones
fragment into several disconnected components. In epidemic modelling, the
largest component of a network corresponds to the maximum number of individuals
who can contract an infection during the outbreak of a disease. In this article,
we show that the size of the largest component in regular networks is always
directly proportional to the number of nodes, regardless of the clustering
coefficient.

A recurrent method in epidemiological studies is the application
of moment closure techniques to obtain results that are intended
to hold approximately and to be independent of a specific graph model~\cite{KeeXX,Kir42,Ser06,Tra07}.
Moment closure methods have been frequently applied to many models
of real-life situations, including those mapped on lattices~\cite{DurXX,Sat94,Rho96,Kle97,FilXX,Ell01,Pay09},
where other analytical methods such as mapping to percolation are
also often used~\cite{Gra98,San02}. The results given by closure
approximations suggest the existence of a phase transition in the
size of the largest component of a regular network at a critical
value of the clustering coefficient $c^\ast = (k-2)/(k-1)$: for
$c<c^\ast$, the number of nodes in the largest component grows linearly
with the total number of nodes $N$; conversely, for $c\geq c^\ast$,
the size of the largest component grows sublinearly with $N$. This
would imply the impossibility of an endemic state of a disease for
$c>c^\ast$, as the fraction of nodes in the largest component would
vanish for $N\rightarrow\infty$. Below, we show numerically the
absence of such a transition in the largest component size. This
indicates that the use of moment closure is not warranted for the
study of clustering-driven fragmentation. Instead, we introduce
an analytical approach that explains exactly how clustering affects
the largest component size of the networks.

\begin{figure}
\centering
\includegraphics[width=0.45\textwidth]{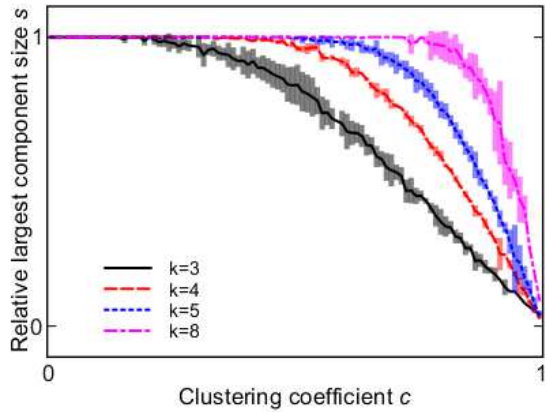}
\caption{\label{Fig1}(Color online) Smooth variation of the largest component size.
The plots of relative largest component size $s$ ($\pm$ one standard deviation)
vs.\ clustering coefficient $c$ show that $s$ does not undergo any phase
transition.}
\end{figure}

To study the dependence of the largest component size on the clustering
coefficient, we performed extensive numerical simulations, generating ensembles
of networks of degree $k=3$, 4, 5 and~8. To create the networks, we used the
direct construction procedure described in~\cite{Del10}. We measured the
clustering coefficient of each network generated,
and computed the relative largest component size
\begin{equation*}
s = \frac{S}{N}\:,
\end{equation*}
where $S$ is the number of nodes found in the largest component. The results,
in Fig.~\ref{Fig1}, show that $s$ varies smoothly with $c$ in all cases. It is
important to note that the degree-based graph construction used guarantees that
means and standard deviations shown represent a combinatorially accurate weighting
over all $k$-regular graphs of a given $c$, without biases that may be induced
by a given random graph model~\cite{Del10,Kim12}.

\begin{figure}
\centering
\includegraphics[width=0.45\textwidth]{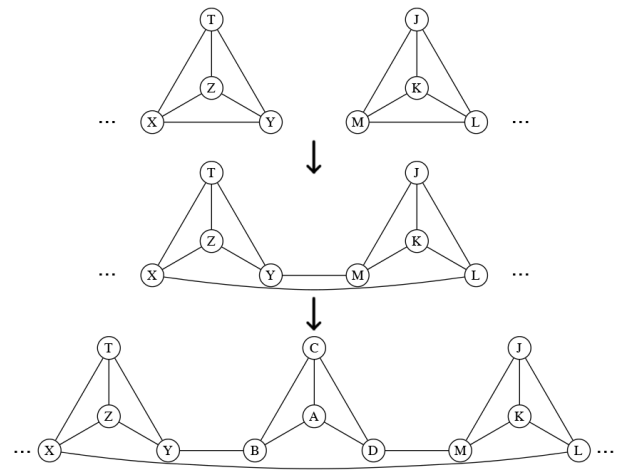}
\caption{\label{Fig2}Example of clique joining for $k=3$.
When $c=1$ the network consists entirely of isolated cliques
(top row). The next highest value of $c$ less than 1 corresponds
to networks in which only two cliques have been joined by
rewiring two pairs of nodes (middle row). The procedure can
be repeated to find networks with the next highest value of
$c$ (bottom row).}
\end{figure}

To understand the absence of a transition, we look at the emergence of a component
of size proportional to $N$ (giant component) as a percolative process. For $N\gg 1$
and $\varepsilon\ll 1$, the ensemble of networks corresponding to $c=1-\varepsilon$
can be generated as follows: start from a network with clustering coefficient $c=1$,
consisting entirely of disconnected cliques of size $k+1$. Then, systematically consider
all pairs of cliques. With probability $\varepsilon$, pick two nodes X and Y from
the first clique, and two nodes L and M from the second, erase the links X--Y and
L--M, and create the links X--L and Y--M (see Fig.~\ref{Fig2}). These new connections establish
``external'' links between local clustered neighbourhoods that were formerly isolated.
Then, we can use the Molloy-Reed criterion~\cite{Mol95} to determine the existence conditions
of a giant component. Applied to our case, the criterion states that a giant component
exists if $\Sigma\equiv\left\langle\sigma^2\right\rangle-2\left\langle\sigma\right\rangle>0$,
where $\sigma$ is the number of external links of a local neighbourhood.  As $\varepsilon$
is small, the probability $P(0)$ of finding a neighbourhood with $\sigma=0$ is of
order $O(1)$. Similarly, $P(2)=O(\varepsilon)$ and $P(4)=O(\varepsilon^2)$. Notice
that $P(1)=P(3)=0$, as every connection between two neighbourhoods requires the rewiring
of two nodes. Thus $\Sigma=O(\varepsilon^2)>0$ and a giant component is always found
for any value of $c$.

To find a functional form for $s(c)$, note that the clustering
coefficient $c$ is the probability that any two nodes which share
a common neighbour are linked. Then, if we were to build a network
by randomly linking its nodes, the probability for a node and
its $k$ neighbours to form a clique of size $k+1$ would be
\begin{equation*}
p = c^\frac{k(k-1)}{2}\:.
\end{equation*}
To find the total number of cliques $n_c$, multiply $p$ by $N$
and divide by $k+1$, to account for the fact that the neighbours
of all the nodes in a clique are linked amongst each other:
\begin{equation*}
n_c = \frac{N}{k+1}c^\frac{k(k-1)}{2}\:.
\end{equation*}
As each clique decreases the fraction of nodes in the largest
component by $(k+1)/N$, it is
\begin{equation}\label{upbou}
 s\left(c\right) = 1-c^\frac{k\left(k-1\right)}{2}\:.
\end{equation}
Then, Eq.~\ref{upbou} provides a simple analytical expression
describing the behaviour of $s$ when $c$ is small enough.

A complete description of $s(c)$ can be obtained using combinatorial
arguments. First, it is straightforward to see that, for $k\geq 3$,
\begin{align*}
 s\left(0\right) &= 1\:,\\
 s\left(1\right) &= 0
\end{align*}
and
\begin{equation}\label{lowder}
 {\left.\frac{\mathrm ds}{\mathrm dc}\right|}_{c=0^+}= 0\:.
\end{equation}
\begin{figure}
\centering
\includegraphics[width=0.45\textwidth]{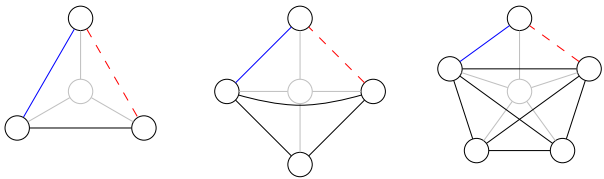}
\caption{\label{Fig3}(Color online) Keeping a leaf node free from triangles with other leaves, for $k=3$, 4 and~5
(from left to right). Each local neighbourhood consists of a central hub node connected
to $k$ leaves (grey links). To have the most links while avoiding triangles between the
top node and other leaves, place all the possible links between the other leaves (black
links), then add one link with the top node (blue link). If any other link is added, for
instance the red one, the top node is no longer free from triangles with other leaves.}
\end{figure}
\begin{figure}
\centering
\vspace*{-3pt}
{\includegraphics[width=0.41\textwidth]{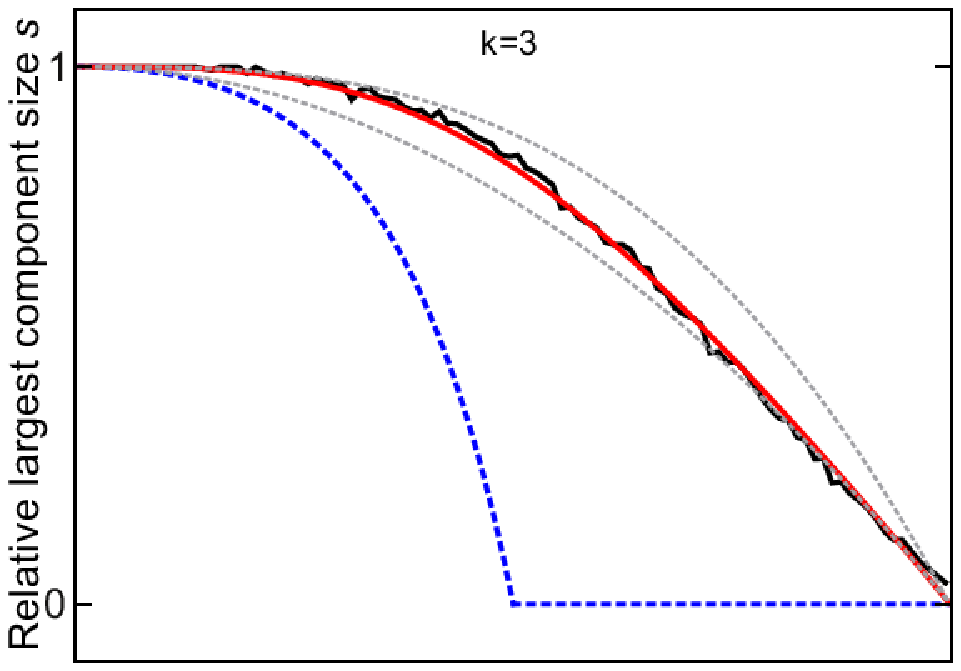}
\vspace*{-7pt}}
{\includegraphics[width=0.41\textwidth]{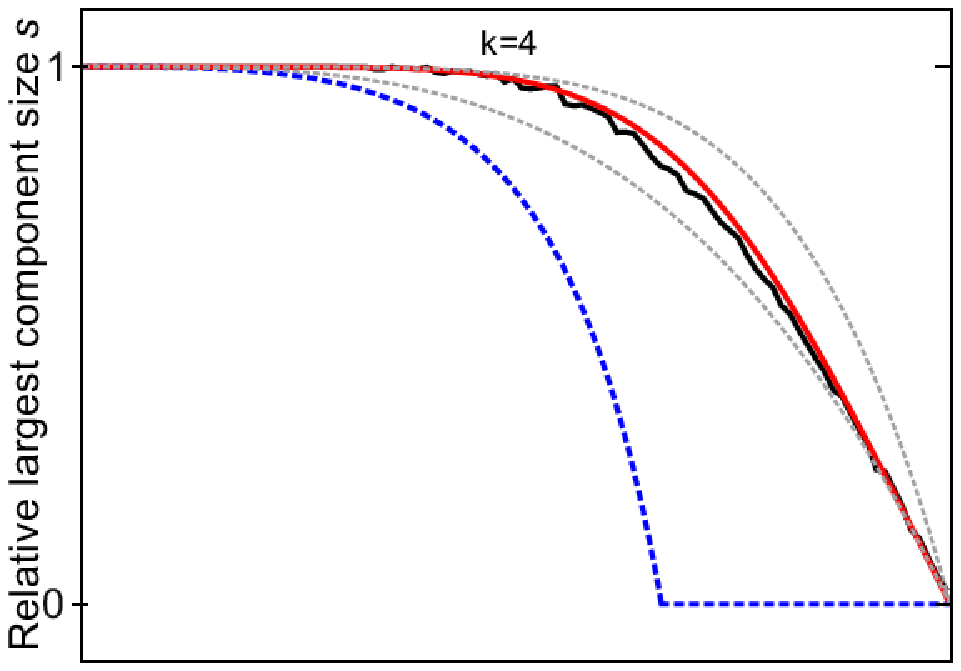}
\vspace*{-7pt}}
{\includegraphics[width=0.41\textwidth]{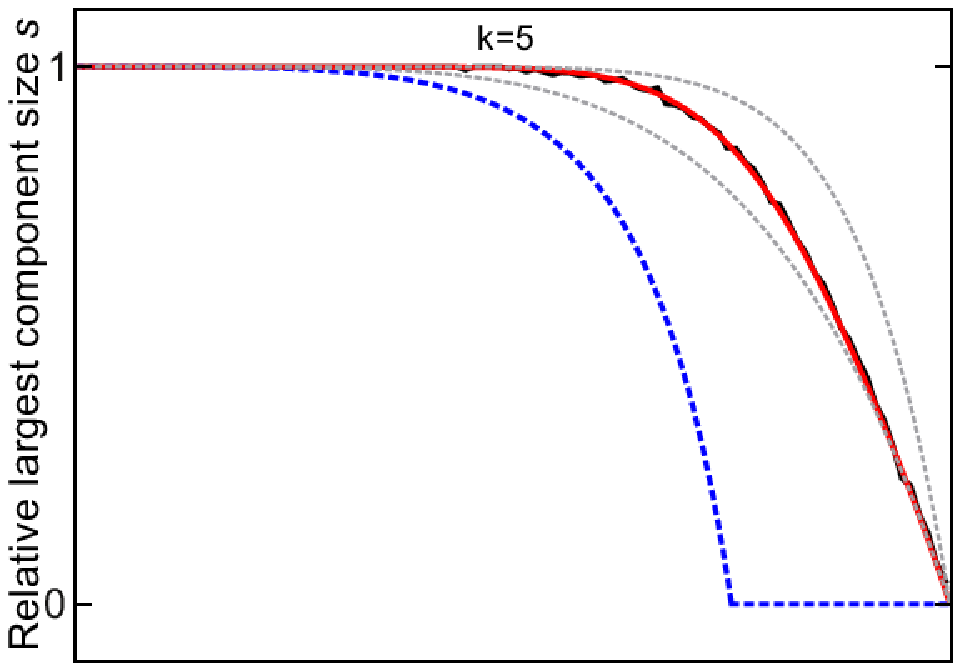}
\vspace*{-7pt}}
{\includegraphics[width=0.41\textwidth]{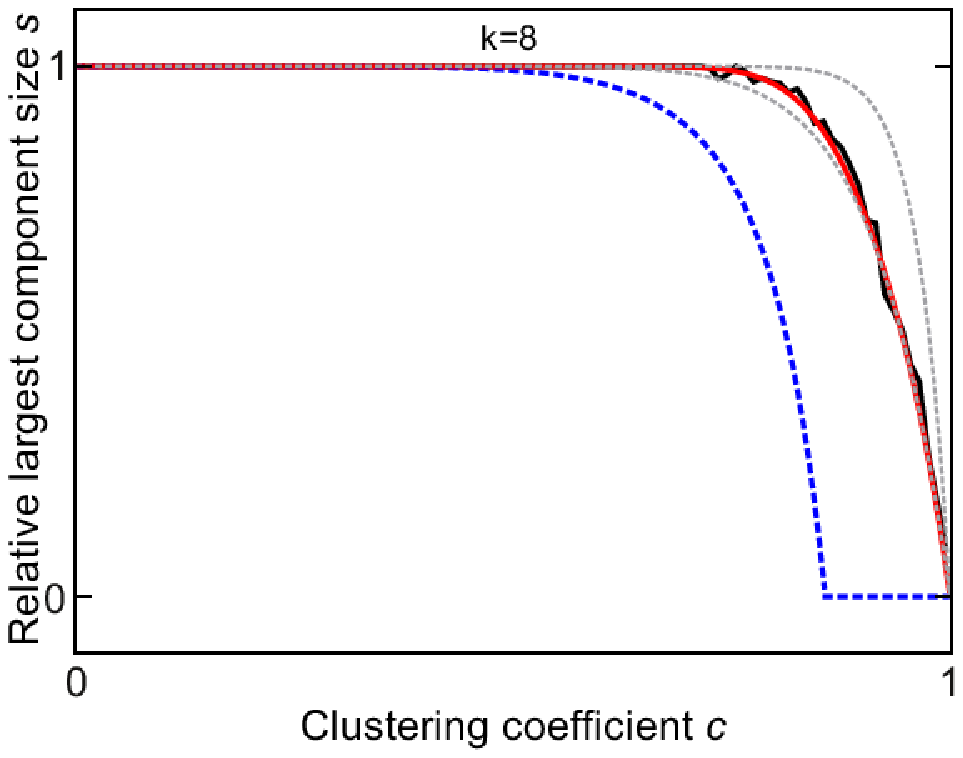}
\vspace*{-15pt}}
\caption{\label{Fig4}(Color online) Permanence of giant component in regular networks.
The plots of relative largest component $s$ vs.\ clustering $c$ show that $s$
always remains directly proportional to the size of the network $N$. The
numerical results (solid black line) are shown together with the prediction
based on moment closure (dashed blue line), the analytical solution (solid
red line) and the asymptotic regimes (dotted grey lines).}
\end{figure}
Then, we can estimate the derivative of $s$ close to $c=1$ using
the ratio of the small finite changes in $s$ and $c$. To do so,
consider a network with the highest possible value of $c<1$. This
can be obtained by starting from a network with $c=1$ and joining
only two cliques using the rewiring described above and in Fig.~\ref{Fig2}.
To construct a network with the next possible highest value of the
clustering coefficient, repeat the rewiring, adding a third clique
to the group. Every time a new clique is added, decreasing $c$,
$k+1$ nodes join the largest component. Thus, the change $\Delta s=-(k+1)/N$.
To compute $\Delta c$, we express $c$ as the mean of the local clustering
coefficients, defined for each node as the number of triangles it
belongs to divided by the maximum number of possible links amongst
its neighbours, which in a regular network is $k(k-1)/2$. Then,
notice that the rewiring procedure breaks 1~triangle for each node
that does not take part in the rewiring. The rewired nodes, instead,
lose an ``internal'' link, and so are left with only $k-1$ neighbours
that form triangles. Thus, if $i$ cliques are joined, then there
are $i(k-1)$ nodes with $k(k-1)/2-1$ triangles each, and $2i$ nodes
with $(k-1)(k-2)/2$ triangles each. Therefore, the contribution to
$c$ coming from the joined cliques is
\begin{multline}\label{clicon}
 c_1 = \frac{2i}{Nk\left(k-1\right)}\left\lbrace\left(k-1\right)\left[\frac{k\left(k-1\right)}{2}-1\right]\right.\\
 \left.+ 2\frac{\left(k-1\right)\left(k-2\right)}{2}\right\rbrace\:.
\end{multline}
Also, the remaining $N-i(k+1)$ nodes in the networks are still
maximally clustered, and so they contribute
\begin{equation}\label{restcon}
 c_2 = 1-\frac{i\left(k+1\right)}{N}\:.
\end{equation}
Summing Eqs.~\ref{clicon} and~\ref{restcon} gives
\begin{equation}\label{highc}
 c = 1-\frac{6i}{kN}\:.
\end{equation}
Eq.~\ref{highc} implies that every time a clique is added, $c$ decreases
of the same amount $\Delta c=6/(kN)$. Therefore, we have
\begin{equation}\label{highder}
 {\left.\frac{\mathrm ds}{\mathrm{d}c}\right|}_{c=1^-}= -\frac{k(k+1)}{6}\:.
\end{equation}
Eqs.~\ref{lowder} and~\ref{highder} show that, for high values of $c$, it is
\begin{equation}\label{lowbou}
  s\left(c\right) = 1-c^\frac{k\left(k+1\right)}{6}\:.
\end{equation}
These results indicate the existence of two regimes,
one for low and one for high values of $c$, given by Eq.~\ref{upbou} and Eq.~\ref{lowbou},
respectively. Then, the general form of $s(c)$ is given by the crossover
behaviour between these two regimes.

To obtain an explicit form for the crossover, we first note that every regular
network is locally a $k$-star, that is, a hub node (or `ego' in sociological
terminology) connected to $k$ outer leaves (or `alters').  We argue that such a
local neighbourhood is highly clustered if it has enough links to guarantee
that each leaf node belongs to at least one triangle made entirely of leaves.
To find how many links are needed to satisfy this constraint, consider the
leaves of a $k$-star and compute the maximum number of links that can be placed
while keeping one leaf triangle-free. This is found by placing all possible
links between $k-1$ leaves, and then adding one further single link~(Fig.~\ref{Fig3}).
At the end of this procedure, one leaf is left with only a single link to other
leaves. Thus, it does not participate in any triangle with the other leaves,
condition that would be broken by the addition of even just one other connection. So,
for a local neighbourhood to be little clustered, the links between its leaves
can be at most $M\equiv (k-1)(k-2)/2+1$. As each link between two leaves exists with probability
$c$, then the probability $W$ that a neighbourhood is not highly clustered is
\begin{equation*}
 W\left(k,c\right) = \sum_{x=0}^M \mathrm{Binom}\left(x|\frac{k\left(k-1\right)}{2},c\right)\:,
\end{equation*}
where $\mathrm{Binom}\left(x|y,z\right)$ is the binomial distribution
for $x$ successes over $y$ trials with probability $z$. The equation
above can be rewritten as~\cite{Fel68}
\begin{equation}\label{weight}
 W\left(k,c\right) = 1-I_c\left(\frac{\left(k-1\right)\left(k-2\right)}{2}+2,k-2\right)\:,
\end{equation}
where $I_t(\alpha,\beta)$ is the regularized incomplete beta function
\begin{equation*}
 I_t\left(\alpha,\beta\right) = \frac{\Gamma\left(\alpha+\beta\right)}{\Gamma\left(\alpha\right)\Gamma\left(\beta\right)}\int_0^t s^{\alpha-1}\left(1-s\right)^{\beta-1}\mathrm ds\:.
\end{equation*}
We can now express the full crossover behaviour of $s(c)$ as an average
of Eqs.~\ref{upbou} and~\ref{lowbou}, weighted with the correct probabilities given by Eq.~\ref{weight}:
\begin{equation}\label{solution}
 s(c) = 1 - c^\frac{k\left(k+1\right)}{6} +
 W\left(k,c\right)\left[c^\frac{k\left(k+1\right)}{6}-c^\frac{k\left(k-1\right)}{2}\right]\:.
\end{equation}
A comparison with the numerical results shows that the
analytical solution closely matches the simulated values,
confirming the correctness of our approach~(Fig.~\ref{Fig4}).

In conclusion, we have demonstrated that the size of the largest component in
regular networks changes smoothly with the clustering coefficient $c$, and
always remains directly proportional to the network size. Thus, regular
networks always have a giant component, even for large values of $c$. We stress
that this novel result is the direct consequence of intrinsic structural
properties of the networks, and it is not based on a particular approximation
used for the calculations. Also, it is to be noted that the conclusions hold
regardless of the transmissivity of a particular disease. Thus, no structural
factor precludes disease transmission, and an endemic state in an epidemic is
therefore always possible in regular networks, as the maximum fraction of
infected individuals does not
vanish in the limit of a large population. Many populations of significant
economic and scientific interest consist of individuals who are sessile or have
a limited epidemiological connectivity. The network topology mostly used in
their study is that of regular graphs. Thus, our results offer new insight into
current epidemiological questions, such as whether structural agricultural
factors can explain the absence of a recent foot-and-mouth outbreak in the
USA~\cite{Til10}. In particular, they suggest that enhanced control and strong
active measures should be undertaken to prevent the spread of a disease to a
substantial part of the population even when the infected system is highly
clustered.

\begin{acknowledgments}
The authors gratefully acknowledge Veselina Uzunova and Lorenzo Pellis
for fruitful discussions. TH is supported by the UK Engineering and Physical
Sciences Research Council. CIDG acknowledges support by EINS, Network
of Excellence in Internet Science, via the European Commission's FP7 under
Communications Networks, Content and Technologies, grant No.~288021.
\end{acknowledgments}


\begin{thebibliography}{99}
\bibitem{Alb02} R.\ Albert and A.-L.\ Barabási, Rev.~Mod.~Phys.\ \textbf{74}, 47 (2002).
\bibitem{Boc06} S.\ Boccaletti et al.\ Phys.\ Rep.\ \textbf{424}, 175 (2006).
\bibitem{Dan11} L.\ Danon et al. Interdiscip.\ Perspect.\ Infect.\ Dis.\ \textbf{2011}, 1 (2011).
\bibitem{Pas01} R.\ Pastor-Satorras and A.\ Vespignani, Phys.\ Rev.\ Lett.\ \textbf{86}, 3200 (2001).  
\bibitem{Shi05} M.~D.~F.\ Shirley and S.~P.\ Rushton, Ecol.\ Complex.\ \textbf{2}, 287 (2005).
\bibitem{Lel09} M.\ Lelarge, in \textit{SIGMETRICS '09: Proceedings of the eleventh international joint conference on Measurement and modeling of computer systems}, Seattle, 2009.
\bibitem{DurXX} S.~A.\ Levin and R.\ Durrett, Phil.\ Trans.\ R.\ Soc.\ Lond.~B \textbf{351}, 1615 (1996); R.\ Durrett and D.\ Remenik, Ann.\ Appl.\ Probab.\ \textbf{19}, 1656 (2009); R.\ Durrett, P.\ Natl.\ Acad.\ Sci.\ USA \textbf{107}, 4491 (2010).
\bibitem{KeeXX} M.~J.\ Keeling, P.\ Roy.\ Soc.~B - Biol.\ Sci.\ \textbf{266}, 859 (1999); T.\ House and M.~J.\ Keeling, J.\ Theor.\ Biol.\ \textbf{272}, 1 (2011).
\bibitem{Giv11} O.\ Givan, N.\ Schwartz, A.\ Cygelberg and L.\ Stone, J.~Theor.\ Biol.\ \textbf{288}, 21 (2011).
\bibitem{Abd11} M.~A.\ Abdullah, C.\ Cooper and M.\ Draief, in \textit{Approximation, Randomization, and Combinatorial Optimization. Algorithms and Techniques}, edited by L.~A.\ Goldberg, K.\ Jansen, R.\ Ravi and J.~D.~P.\ Rolim (Springer-Verlag, Berlin, 2011), p.~351.
\bibitem{Kee00} M.~J.~Keeling and C.~A.~Gilligan, Nature \textbf{407}, 903--906 (2000)
\bibitem{Fer01} N.~M.\ Ferguson, C.~A.\ Donnelly and R.~M.\ Anderson, Science \textbf{292}, 1155 (2001).
\bibitem{Gib97} G.~J.\ Gibson, Phytopathology \textbf{87}, 139 (1997).
\bibitem{Del11} C.~I.\ Del~Genio, T.\ Gross and K.~E.\ Bassler, Phys.\ Rev.\ Lett.\ \textbf{107}, 178701 (2011).
\bibitem{Wat98} D.~J.\ Watts and S.~H.\ Strogatz, Nature \textbf{393}, 440 (1998).
\bibitem{NewXX} M.~E.~J.\ Newman, SIAM~Review \textbf{45}, 167 (2003); M.~E.~J.\ Newman, Phys.\ Rev.\ Lett.\ \textbf{103}, 058701 (2009); M.~E.~J.\ Newman, \textit{Networks: an introduction} (Oxford University Press, Oxford, United Kingdom, 2010).
\bibitem{Vol04} E.\ Volz, Phys.\ Rev.~E \textbf{70}, 056115 (2004).
\bibitem{Gle09} J.~P.\ Gleeson and S.\ Melnik, Phys.\ Rev.~E \textbf{80}, 046121 (2009).
\bibitem{Bol11} B.\ Bollobás, S.\ Janson, and O.\ Riordan, Random.\ Struct.\ Algor.\ \textbf{38}, 269 (2011).
\bibitem{BriXX} T.\ Britton, M.\ Deijfen, A.~N.\ Lagerås and M.\ Lindholm, J.~Appl.\ Probab.\ \textbf{45}, 743 (2008); F.\ Ball, D.\ Sirl and P.\ Trapman, Math.\ Biosci.\ \textbf{224}, 53 (2010); F.\ Ball, T.\ Britton and D.\ Sirl, J.~Math.\ Biol.\ \textbf{66}, 979 (2013).
\bibitem{Mos09} M.\ Moslonka-Lefebvre, M.\ Pautasso, M.~J.\ Jeger, J.~Theor.\ Biol.\ \textbf{260}, 402 (2009).
\bibitem{Mol12} C.\ Molina and L.\ Stone, J.~Theor.\ Biol.\ \textbf{315}, 110 (2012).
\bibitem{Kir42} J.~G.\ Kirkwood and E.~M.\ Boggs, J.\ Chem.\ Phys.\ \textbf{10}, 394 (1942).
\bibitem{Ser06} M.~A.\ Serrano and M.\ Boguñá, Phys.\ Rev.\ Lett.\ \textbf{97}, 088701 (2006).
\bibitem{Tra07} P.\ Trapman, Math.\ Biosci.\ \textbf{210}, 464 (2007).
\bibitem{Sat94} K.\ Satō, H.\ Matsuda and A.\ Sasaki, J.~Math.\ Biol.\ \textbf{32}, 251 (1994).
\bibitem{Rho96} C.~J.\ Rhodes and R.~M.\ Anderson, J.~Theor.\ Biol.\ \textbf{180}, 125 (1996).
\bibitem{Kle97} A.\ Kleczkowski, C.~A.\ Gilligan and D.~J.\ Bailey, Proc.\ R.\ Soc.\ Lond.~B \textbf{264}, 979 (1997).
\bibitem{FilXX} J.~A.~N.\ Filipe and G.~J.\ Gibson, Phil.\ Trans.\ R.\ Soc.\ Lond.~B \textbf{353}, 2153 (1998); J.~A.~N.\ Filipe and M.~M.\ Maule, Math.\ Biosci.\ \textbf{183}, 15 (2003).
\bibitem{Ell01} S.~P.\ Ellner, J.~Theor.\ Biol.\ \textbf{210}, 435 (2001).
\bibitem{Pay09} J.~L.\ Payne and M.~J.\ Eppstein, Evol.\ Comput.\ \textbf{17}, 203 (2009).
\bibitem{Gra98} P.\ Grassberger, Math.\ Biosci.\ \textbf{63}, 157 (1998).
\bibitem{San02} L.~M.\ Sander et al.\ Math.\ Biosci.\ \textbf{180}, 293 (2002).
\bibitem{Del10} C.~I.\ Del~Genio, H.\ Kim, Z.\ Toroczkai, and K.~E.\ Bassler, PLoS ONE \textbf{5}(4), e10012 (2010).
\bibitem{Kim12} H.\ Kim, C.~I.\ Del~Genio, K.~E.\ Bassler and Z.\ Toroczkai, New J.~Phys.\ \textbf{14}, 023012 (2012).
\bibitem{Mol95} M.\ Molloy and B.\ Reed, Random Struct.\ Algor.\textbf{6}, 161 (1995).
\bibitem{Fel68} W.\ Feller, \textit{An introduction to probability theory and its applications} (Wiley, New York, USA, 1968).
\bibitem{Til10} M.~Tidlesley, T.~House, M.~Bruhn, R.~Curry, M.~O'Neill, G.~Smith and M.~J.~Keeling, PNAS \textbf{107}, 3 (2010).
\end{thebibliography}
\end{document}